\documentclass[a4paper, twocolumn]{article}

\usepackage{siunitx}
\usepackage{xcolor}
\usepackage{physics}
\usepackage{authblk}
\usepackage{graphicx}
\usepackage{hyperref}
\usepackage[style=numeric, sorting=none, hyperref,
    url=false,
	doi=false,
	isbn=false,
	eprint=false,
	giveninits=true,
            sortcites=true,
            sorting=none,
            language=british]{biblatex}

\newbibmacro{string+doi}[1]{%
  \iffieldundef{doi}{%
    \iffieldundef{url}{#1}{\href{\thefield{url}}{#1}}
    }{%
    \href{http://dx.doi.org/\thefield{doi}}{#1}%
  }}
\DeclareFieldFormat{title}{\usebibmacro{string+doi}{\mkbibemph{#1}}}
\DeclareFieldFormat[article]{title}{\usebibmacro{string+doi}{\mkbibquote{#1}}}
\DeclareFieldFormat[thesis]{title}{\usebibmacro{string+doi}{\mkbibquote{#1}}}

\AtEveryBibitem{%
  \clearfield{note}%
	}

\definecolor{lapis}{HTML}{22577A}
\definecolor{cosmos}{HTML}{650D1B}
\definecolor{orange}{HTML}{FF6B35}
\definecolor{asparagus}{HTML}{679436}
\definecolor{grey}{HTML}{4A5759}

\definecolor{pink}{HTML}{D73665}

\newcommand{\myeqnref}[1]{Eqn.~\ref{#1}}
\newcommand{\myfigref}[1]{Fig.~\ref{#1}}

\newcommand{\mysubfigref}[2]{Fig.~\hyperref[#1]{\ref*{#1}~(#2)}}
\newcommand{\mysubfigureref}[2]{Subfigure.~\hyperref[#1]{\ref*{#1}~(#2)}}
\newcommand{\inlineref}[1]{Ref.~\cite{#1}}
\newcommand{\VGIF}{\ensuremath{V_\text{G}}}
\newcommand{\LHCF}{\ensuremath{L_\text{H}}}

\newcommand{\esRb}{\textsuperscript{87}Rb}
\newcommand{\rG}{\ensuremath{\rho_\text{G}}} 
\newcommand{\rH}{\ensuremath{\rho_\text{H}}} 
\newcommand{\rS}{\ensuremath{\rho_\text{S}}} 
\newcommand{\psiGIF}{\ensuremath{\psi_\text{G}}}
\newcommand{\psiHCF}{\ensuremath{\psi_\text{H}}}
\newcommand{\psiSMF}{\ensuremath{\psi_\text{S}}}

\author[1,*]{Cameron McGarry}
\author[1]{Kerrianne Harrington}
\author[1]{Daniel J. Goodwin}
\author[1]{Charles Perek-Jennings}
\author[1]{Tim A.\ Birks}
\author[1]{Kristina R.\ Rusimova}
\author[1]{Peter J.\ Mosley}

\affil[1]{Centre for Photonics and Photonic Materials, Department of Physics, University of Bath, Bath, BA2 7AY, UK}
\affil[*]{cdm34@bath.ac.uk}

\begin{document}

\title{Low-loss, compact, fibre-integrated cell for quantum memories}

\maketitle

\textbf{
We present a low-loss, compact, hollow core optical fibre (HCF) cell integrated with single mode fibre (SMF).
The cell is designed to be filled with atomic vapour and used as a component in photonic quantum technologies, with applications in quantum memory and optical switching.
We achieve a total insertion loss of \SI{0.6+-0.2}{\decibel} at \SI{780}{\nano\meter} wavelength via graded index fibre to ensure efficient mode matching coupled with anti-reflection coatings to minimise loss at the SMF-HCF interfaces.
We also present numerical modelling of these interfaces, which can be undertaken efficiently without the need for finite element simulation.
We encapsulate the HCF core by coupling to the SMF inside a support capillary, enhancing durability and facilitating seamless integration into existing fibre platforms.
}


\section{Introduction}

Quantum technologies promise to utilise the fundamental principles of quantum mechanics to deliver unparalleled possibilities for enhanced data security through quantum cryptography~\cite{Gisin2002}, increased computational power via quantum computing~\cite{Bennett2000}, and ultra-precise measurement with quantum sensing~\cite{Degen2017}.
The high bandwidth and low decoherence of photons make them a promising candidate as an architecture for quantum technologies~\cite{OBrien2007}, especially in quantum key distribution~\cite{Gisin2007} and distributing entanglement between nodes in quantum networks~\cite{Kimble2008}, both of which can be achieved at long range by means of conventional optical fibres.

However, challenges remain in achieving high bandwidth, low-loss storage and multiplexing of photonic quantum states required for such schemes~\cite{FrancisJones2016, Makino2016}.
The development of optical quantum memories~\cite{Finkelstein2018, Thomas2023} and optical switching~\cite{Davis2023} have revealed a pathway to overcoming these challenges. These demonstrations have predominantly been in macroscopic free space components or on photonic integrated circuits\cite{Mottola2023}, both of which result in high losses when interfaced with optical fibres. This can destroy quantum correlations and negate any quantum advantage.

One route to mitigating this loss is the integration of materials for storage and switching -- for example atomic vapours -- directly with optical fibre~\cite{Sprague2014}. 
Previous work has predominantly been focused on interfacing atoms with the evanescent field of optical nanofibres~\cite{Rajasree2020}, which have been used to implement both quantum memories~\cite{Gouraud2015} and phase modulation~\cite{Vetsch2010}.
Using such a device, enhancement is possible by the implementation of an in-fibre cavity~\cite{Kato2015} or by introducing an external microresonator~\cite{OShea2013, Will2021}. Similar experiments have been performed for atoms trapped near on-chip photonic crystal cavities~\cite{Thompson2013, Tiecke2014}.

An alternative to coupling atoms into the evanescent field is to directly couple to the mode propagating within the fibre.
This allows a greater overlap of the light and the atoms, resulting in a stronger interaction between the two.
To achieve this requires a fibre with a hollow core.
Loading of atoms into such fibres has been demonstrated both with warm vapours by diffusion~\cite{Haupl2022} and with laser-cooled atoms by controlled loading~\cite{Bajcsy2011, Peters2021}.
Once loaded, atoms can be controlled to implement protocols in fibre, such as interferometry~\cite{Xin2018}, all-optical switching~\cite{Peyronel2012} or the generation of Rydberg atoms~\cite{Epple2014, Menchetti2023} which can then be addressed by radio frequency fields, with potential applications as elctro-optic components~\cite{Veit2016}. 
Although this previous work demonstrates the potential of fibre-integrated atomic experiments, they are inherently difficult to scale due to the nature of the vacuum systems required.
Even when we have a warm atomic gas in a fibre (as in \inlineref{Haupl2022}), we run into a second roadblock: the high coupling losses that occur at the fibre interfaces will inhibit the implementation of robust quantum protocols.

The realisation of scalable atomic memory schemes in fibre-integrated systems therefore has a number of requirements.
First, the atoms should be encapsulated sufficiently near to, or within, the optical fibre, without the need for supporting vacuum equipment.
Second, light should be able to traverse through the atoms with sufficiently low loss so as to be compatible with the photonic quantum computing schemes~\cite{Pankovich2023}.
The low-loss wavelength must match that at which the chosen quantum protocol operates.
And finally, the rate at which the atoms collide with the fibre should be small compared to the lifetime of the memory: this is to avoid decoherence of the atomic states upon collision with the walls.
That is to say, the core of the fibre should be sufficiently large as to not significantly limit the memory lifetime.

\begin{figure*}[htbp]
  \centering
  \includegraphics[width=130mm]{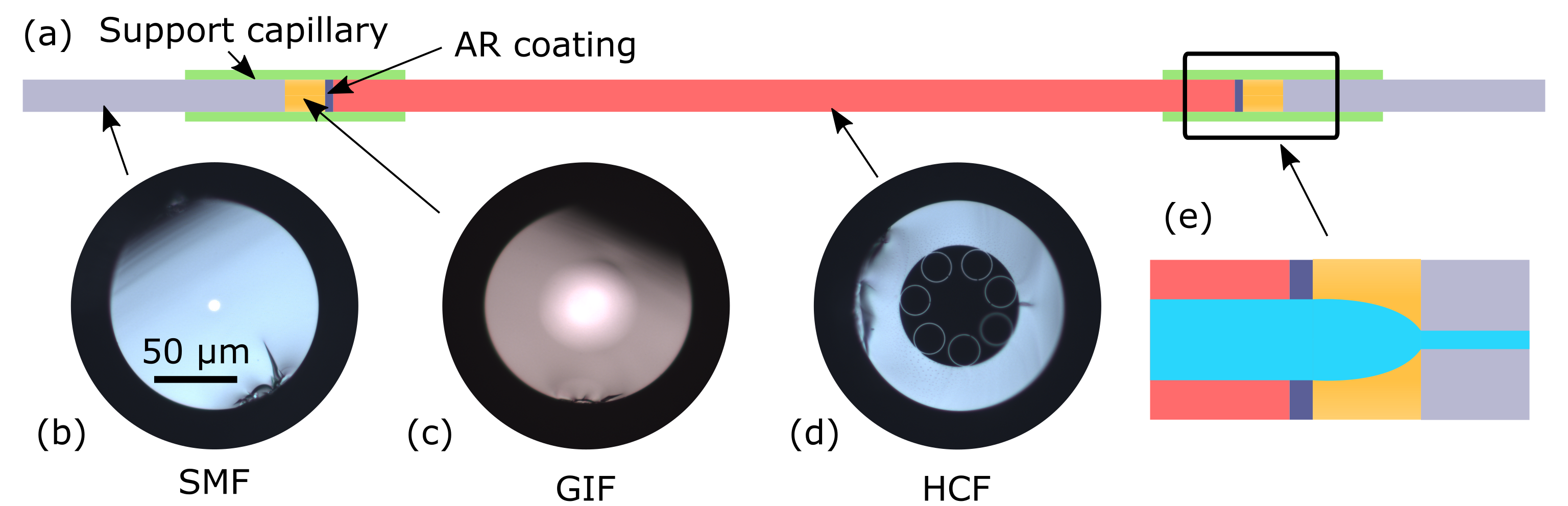}
\caption{(a) Schematic of the SMF-GIF-HCF vapour cell. (b) Optical micrograph of the SMF, with the core illuminated. (c) Optical micrograph of the GIF with the core illuminated (same scale). (d) Optical micrograph of the fabricated HCF (same scale). (e) Schematic of mode expansion by GIF lens (light shown in blue). }
\label{scheme}
\end{figure*}

Here, we realise a low-loss, fibre-integrated cell which, fulfills the above criteria and, once filled with an atomic vapour, can be used to realise an in-fibre quantum memory.
Our device is pictured in \myfigref{scheme}.
The first criterion is satisfied by use of a negative-curvature anti-resonant HCF, in which light is confined to a central hollow core by tubular resonators~\cite{Yu2016}.
This will allow for the loading of atomic vapours within the core, where they will be strongly coupled to the propagating light field.
Encapsulation of the core is achieved by mounting the ends of the HCF in a support capillary, containing an interconnection to SMF.

The second criterion is fulfilled by implementation of mode-matching via a short segment of graded index fibre (GIF)~\cite{Komanec2019}. 
The GIF serves to expand the mode of the SMF to match that of the HCF, and allow for low-loss interconnections.
A schematic of the mode expansion as well as a cross section of all three fibres is shown in \myfigref{scheme}.
Previously, similar interconnects have been used to form vapour cells for inert gases with applications in classical sensing~\cite{Suslov2023}, without demonstrating the low loss required for quantum applications. In contrast, our cell achieves low loss at \SI{780}{\nano\meter} wavelength, which corresponds to the \esRb{} $D_2$ line widely used in a variety of quantum memory techniques~\cite{Finkelstein2018, Thomas2022, Kaczmarek2017}.

The HCF used here (see \mysubfigref{scheme}{d}) has a large core relative to the transmission wavelength of interest, $\lambda = \SI{780}{\nano\meter}$ with radius $\rH=\SI{17}{\micro\meter}$ ($2\rH/\lambda=44$)~\cite{Song2019, Yu2016}.
As we will discuss below, this is principally to enable efficient mode matching, but additionally will help to mitigate any decrease in memory lifetime by collision-induced decoherence.
We note also that the HCF cell offers a particular advantage for optical phase modulation in rubidium, where long spin lifetimes are not required as the switching occurs on the timescale at which light traverses the vapour cell~\cite{Davis2023}.
Further, localising light to the length scale of the HCF core enables high intensities to be maintained over many times the Rayleigh range of a free space beam the size of a guided mode, boosting the efficiency of light-matter interactions.

\section{Numerical methods}

In this section, we will describe a numerical model used to optimse the SMF-GIF-HCF coupling.
Similar calculation is possible by considering the optical mode in a Gaussian ray picture~\cite{Emkey1987} and modelling the GIF segment as a lens. Here however, we will directly consider the propagation of the modes within the fibres. 
The numerical implementation used below is made available in the form of an IPython notebook in \inlineref{Jupyter}.

We note that our cell design is constrained by the decision to use commercially available SMF (Thorlabs 780HP~\cite{780HP}) and GIF (Thorlabs GIF625~\cite{GIF625}); this is convenient for integration with existing fibre infrastructure and for the construction of individual devices, but limits the control of the coupling.
All that can now be varied is the light pattern exiting the GIF (by the GIF length) and the mode shape in the HCF (by its core radius, which is determined when we draw the fibre).
The methods presented here remain generally applicable to the SMF-GIF-HCF interconnects, however numerical values are for this particular choice of fibres.

We begin by considering the coupling efficiency between two fibres, which is given by the overlap integral of the two light patterns (represented by $\psi_1$ and $\psi_2$)~\cite{SandL}
\begin{equation}
    \eta(\psi_1, \psi_2) = \frac{\left|\int \psi_1(\mathbf{r})\psi_2(\mathbf{r})\dd S\right|^2}{%
    \int \left|\psi_1(\mathbf{r})\right|^2\dd S%
    \int \left|\psi_2(\mathbf{r})\right|^2\dd S}.
    \label{overlap}
\end{equation}
where the integral is taken over the entire area of the light patterns at the interface.

The SMF mode has the usual form~\cite{SandL}
\begin{equation}
   \psiSMF(R) = \begin{cases}
   \frac{J_0(\tilde{U}R/\rS)}{J_0({\tilde{U}})}, & R\leq\rS \\[10pt]
   \frac{K_0(\tilde{W}R/\rS)}{K_0({\tilde{W}})}, & R>\rS
   \end{cases}.
\end{equation}
where $R$ is the radial position and $\rS = \SI{2.2}{\micro\meter}$ is the SMF core radius, while $\tilde{U} = 1.636$ and $\tilde{W} = 1.704$ are the relevant core and cladding parameters respectively.
These parameters are calculated numerically by solving the characteristic equation of the waveguide~\cite{SandL}; see \inlineref{Jupyter}.
We also have $J_n$ as the $n^\text{th}$ Bessel function of the first kind and $K_n$ as the $n^\text{th}$ modified Bessel function of the second kind.

We now consider the light pattern in the GIF, which is fully described in \inlineref{SandL}. This light is multimoded, so that light exiting GIF of length $L$ has light pattern given by
\begin{equation}
    \psiGIF(R, L) = \sum_{l,m}a_{l,m}g_{l,m}(R)e^{i\beta_{l,m} L}.
\end{equation}
Here we have introduced the form of the individual modes,
\begin{equation}
  g_{l,m}(R) = R^l \mathcal{L}^{(l)}_{m-1}(\VGIF R^2)\exp\left(-\frac{\VGIF R^2}{2}\right),
\end{equation}
where $\mathcal{L}^{(l)}_{m-1}$ as a Laguerre polynomial and $\VGIF=66.2$ is the GIF fibre parameter.
The propagation constant is
\begin{equation}
    \beta_{l,m} = \frac{\VGIF}{\rG} \sqrt{2\Delta}\left[1-\frac{2\rG^2g^2}{\VGIF}(2m + l -1)\right]^{1/2},
\end{equation}
where $\Delta=1.78\times10^{-2}$ is the profile height parameter of the GIF~\cite{SandL, Jupyter}.
We have introduced the amplitudes of the various modes $a_{l,m}$; these can be calculated from the overlap integral (\myeqnref{overlap}) of the GIF mode at $L=0$ with the mode of the SMF. That is,
\begin{equation}
a_{l,m} \propto \int_0^\infty g_{l,m}(R) \psiSMF(R) R\dd R,
\end{equation}
the constant of proportionality is determined by normalising $\psiGIF$; we assume that the loss will be dominated the GIF-HCF interconnect, and so set $\eta(\psiGIF, \psiGIF) = 1$.
This is trivially verified experimentally by measuring the loss during fusion splicing of the SMF and GIF, which is negligible.
This calculation allows the determination of any value of $\psiGIF(R, L)$, as each term in the $\psiGIF$ expansion can be numerically determined independently from the others.

The transfer efficiency from GIF to HCF is found by considering the overlap integral with the HCF mode.
We assume that this is the fundamental ($\text{LP}_{0,1}$) mode of the HCF, which we take to be circular, reaching zero at the extent of the core~\cite{Murphy2023}. This mode has the form
\begin{equation}
   \psiHCF (R, \rH) = \begin{cases}
   J_l(j_{l,m} R/\rH), & R\leq\rH \\[10pt]
   0 & R > \rH
   \end{cases},
\end{equation}
of an $\text{LP}_{l,m}$ mode with $l-=0$ and $m=1$, where $j_{l,m}$ is the $m^\text{th}$ zero of the Bessel function $J_l$.
Since we assumed that no light is lost at the SMF-GIF interface, the total efficiency of the SMF-GIF-HCF interconnect is given by $\eta(\psiGIF, \psiHCF)$.
As noted above, the system is constrained so that we can vary only the GIF length ($L$) and the HCF core radius ($\rH$).
The coupling efficiency is therefore a function of these two variables, which we rewrite as $\eta = \eta(\psiGIF, \psiHCF) = \eta(L, \rH)$.

\begin{figure*}[htbp]
  \centering
  \includegraphics[width=130mm]{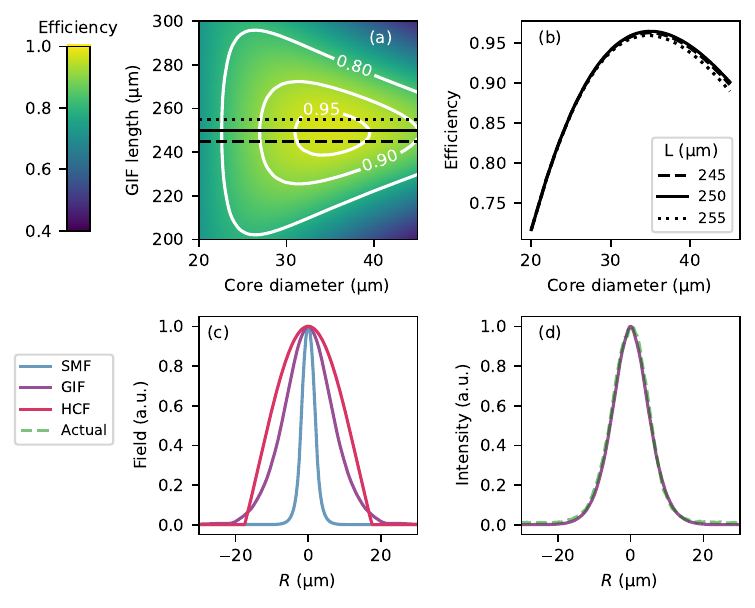}
  \caption{Effect of varying core diameter and GIF length on the coupling efficiency. (a) heat map of the coupling, with contours highlighting various efficiencies. Black lines correspond to cut-through plots shown in (b). (b) Efficiency varying with core size for fixed values of $L$. (c) mode profiles in the SMF, GIF and HCF for optimised coupling with profile heights set arbitrarily. (d) comparison between the calculated and measured intensity of light exiting the GIF.}
\label{couple}
\end{figure*}

The results of the numerical calculations of the coupling efficiency are shown in \myfigref{couple}, where we see that the optimum coupling efficiency occurs for $L=\SI{250}{\micro\meter}$ and $2\rH=\SI{35}{\micro\meter}$, commensurate with results produced by finite element simulations~\cite{Zuba2023}.
Additionally, we measure the intensity of the near field light exiting a GIF (\mysubfigref{couple}{d}), which has excellent agreement with the numerical result.

\section{Fabrication methods}

We will now detail the steps taken to build our device, including the fabrication of custom HCF.
The desired properties of the HCF are determined from both optical and mechanical requirements.
Optically, the HCF must have resonator thickness that will allow transmission at the desired wavelength.
For $\lambda=\SI{780}{\nano\meter}$ we aim for a nominal resonator thickness of \SI{0.56}{\micro\meter}.
Additionally, from the numerical methods in the previous section, we have determined a desired core diameter of \SI{35}{\micro\meter}.
Mechanically, the outer diameter (OD) of the fibre must match that of the SMF and GIF (\SI{125}{\micro\meter}) to allow mating in the capillary.
Custom capillary with nominal internal diameter (ID) \SI{126}{\micro\meter} was also drawn.

We fabricated HCF using the stack and draw method, building a macroscopic stack of seven resonators held by supports before drawing into canes and later fibre.
The seven-resonator design was chosen to yield the desired core diameter and OD with the available glass.
A series of fibre draws were undertaken, with the optimal draw parameters determined experimentally.
An optical micrograph showing the cross section resulting HCF is shown in \mysubfigref{scheme}{d}, alongside the SMF (d) and GIF (e) for comprison.
The HCF transmission properties are shown in \myfigref{loss}, with cut-back measurement demonstrating an attenuation of \SI{0.03+-0.02}{\decibel\per\meter} at \SI{780}{\nano\meter} wavelength, which is typical of HCF in this region of the spectrum~\cite{Yu2016}.
%

We constructed the fibre cell by fusion splicing a section of GIF and SMF.
The GIF was then cleaved using a tension cleaver. We were able to precisely control the cleave position (and hence achieve the desired $L$) by means of a micrometer screw gauge, which we have installed on the tension cleaver.
The cleaved end-faces of these SMF-GIF pigtails were then coated with an anti-reflection (AR) coating by Optical Fibre Packaging Ltd.\ to reduce Fresnsel reflections at the GIF-HCF interface.

Alignment of the SMF-GIF pigtail inside the capillary must be achieved without damaging the fibre end faces.
This was achieved by the following process: first, the HCF was stripped of coating and threaded through the capillary by hand, such that the end of the fibre protruded from the opposing side.
This end of the HCF was then cleaved, ensuring that there was a clean end face, and the capillary was mounted on a three-axis fibre alignment stage.
The SMF pigtail was mounted opposite the HCF, and the two were aligned to optimise the throughput of an alignment laser (ThorLabs S1FC780PM) between the fibres.
This procedure brought the pigtail and the HCF into alignment; by extension the capillary and the pigtail were also aligned.
The HCF was pulled back inside the capillary, leaving both the SMF and the capillary in place.
The SMF was then moved inside the capillary by translating it along the fibre-capillary axis; as the two were already aligned. This allowed the SMF to enter the capillary without damaging the end face of the fibre.
Further adjustment of the SMF's position on the fibre-capillary axis allowed for re-optimisation of the coupling with the alignment laser so as to maximise transmission.
The entire assembly was then secured in place by the application of adhesive to either end, engaging the SMF and the HCF to the capillary and forming a seal at that end of the HCF.
We repeated this procedure at the other end of the HCF to form the encapsulated cell as depicted in \mysubfigref{scheme}{a}.

\section{Device characterisation}

\begin{figure}[htbp]
  \centering
  \includegraphics[width=8cm]{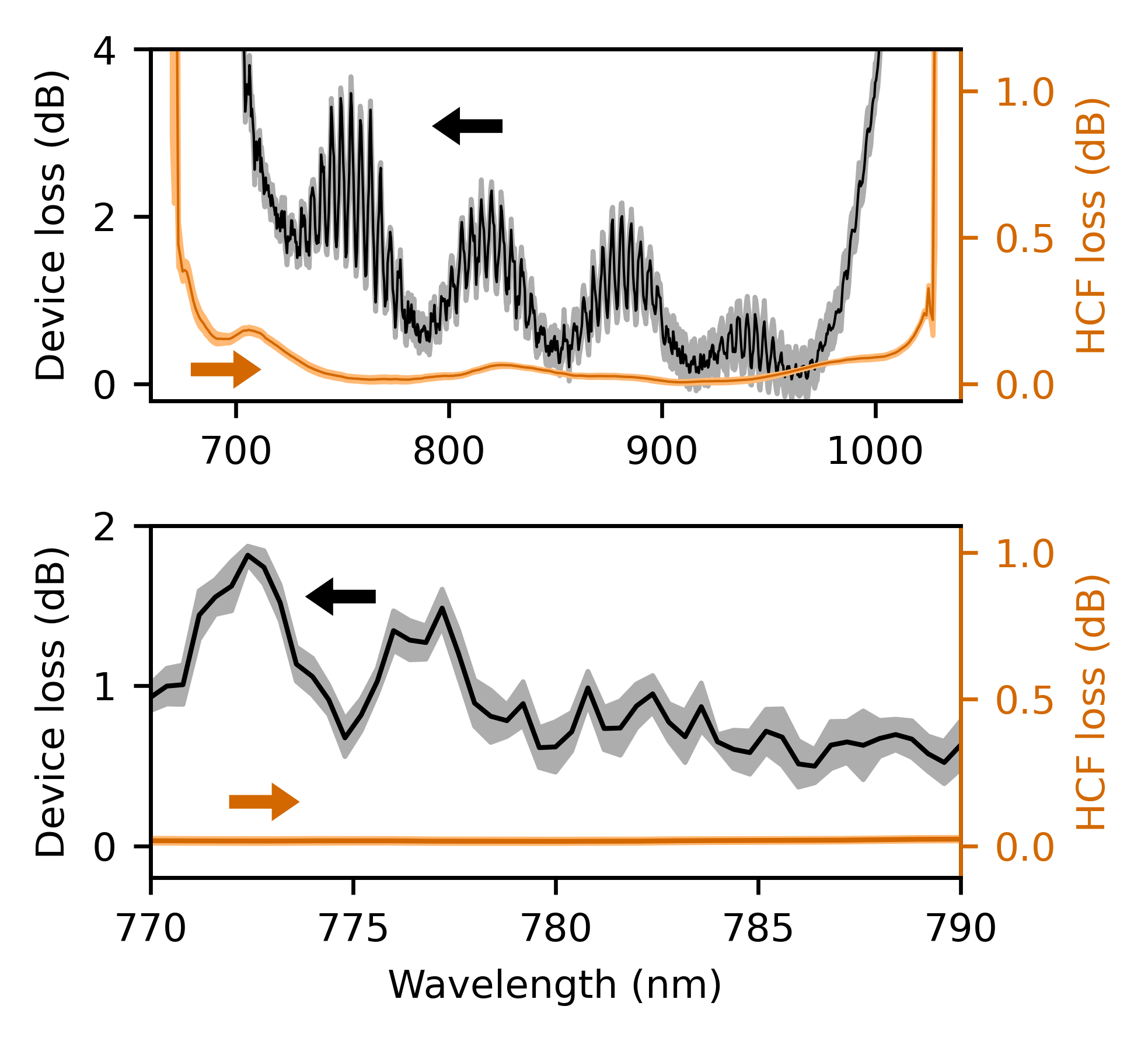}
  \caption{Loss spectrum of the optical fibre cell (black) across a wide wavelength range (top) and zoomed view close to the wavelength of interest (bottom). The loss of the HCF over \SI{50}{\centi\meter} (orange) is shown for comparison. This was determined by a cutback measurement across \SI{18.4}{\meter} of fibre which gave an attenuation of  \SI{0.03+-0.02}{\decibel\per\meter}.}
\label{loss}
\end{figure}

We measured a loss spectrum of the device by the cutback technique, taking a spectrally resolved power measurement at the device output and then cutting back across both interconnects to measure the light in the input SMF.
The results are displayed in \myfigref{loss}, showing a total loss across the entire device of \SI{0.6+-0.2}{\decibel} at \SI{780}{\nano\meter}, dominated by loss at the interconnects. The HCF length used here is \SI{50}{\centi\meter}.

Other regions of the transmission spectrum show lower loss, consistent with the minimum achievable, however we are chiefly concerned with performance at \SI{780}{\nano\meter} for compatibility with rubidium based memory schemes.
We suggest that additional loss arises from a combination of misalignment of the GIF-HCF interface and in-fibre losses.
Although the support capillary is designed to have an ID of \SI{126}{\micro\meter}, the tolerance of this is not fully characterised, and can vary along the length by several microns.
The effect of nonconcentricity of the two fibres (and hence the two modes) in the support capillary can be investigated by considering the overlap integral of spatially separated modes (by \myeqnref{overlap}), and can be shown to have a significant effect (a \SI{2}{\micro\meter} offset can cause a \SI{2.4}{\percent} decrease in efficiency).
%

\begin{figure}[htbp]
  \centering
  \includegraphics[width=8cm]{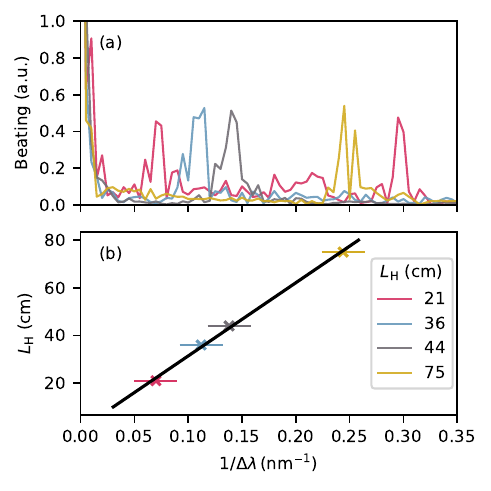}
\caption{(a) Fourier analysis of beating for various fibre lengths $\LHCF$. (b) Linear relationship between the beat spacing $\Delta\lambda^{-1}$ and $\LHCF$.}
\label{phase_length}
\end{figure}

We attribute the rapid beating in the transmission spectrum to higher order modes (HOMs) propagating in the fibre.
This is confirmed by the construction of a series of fibre cells with various lengths of HCF ($\LHCF$).
The phase difference between two modes propagating through the HCF at the second HCF-GIF interface will be $\phi=(\beta_a - \beta_b)\LHCF$, where $\beta_i$ is the propagation constant of mode $i\in\{a,b\}$.
The spacing of the peaks can be found by differentiation, so that
\begin{equation}
    \frac{\dd \phi}{\dd \omega} = \frac{2\pi}{\Delta\omega} = \left(\frac{\dd\beta_a}{\dd\omega} - \frac{\dd\beta_b}{\dd \omega}\right)\LHCF.
\end{equation}
We relate the spacing in frequency to that in wavelength by $\Delta\lambda = \lambda^2\Delta\omega/(2\pi c)$, and the propagation constant to the fibre core parameter of the mode by $U_i = \rH\sqrt{(kn)^2 - \beta_i^2}$.
Here we have $n=1$ as the refractive index of the HCF air core.
Noting also that 
\begin{equation}
    \frac{\dd\beta_i}{\dd\omega} = \frac{1}{c}\frac{\dd \beta_i}{\dd k} \approx kn - \frac{U_i^2}{2kn\rH^2},
\end{equation}
we can relate the beat length in wavelength space to the HCF length
\begin{equation}
    \frac{1}{\Delta\lambda} = \frac{U_b^2 - U_a^2}{8\pi^2\rH^2}\LHCF.
    \label{beating}
\end{equation}
For an HCF, $U_a$ and $U_b$ correspond to $j_{l,m}$ for the two modes that couple.Hence assuming that all beating occurs with mode $a$ as the fundamental ($U_a = j_{0,1} = 2.405$), it is possible to determine the HOM(s) from $\Delta\lambda$ and $\LHCF$.
This is shown in \myfigref{phase_length}, where we present the Fourier Transform of transmission data of devices with various $\LHCF$.
A fit of $\Delta\lambda^{-1}$ against $\LHCF$ shows that $U_b=\SI{3.63+-0.16}{}$, which suggests that beating is between the fundamental mode and the $\text{LP}_{1,1}$ mode ($j_{1,1} = 3.83$).
Note that for short fibre lengths it is possible to resolve beating into additional HOMs, as can be seen for the additional peak at \SI{0.3}{\per\nano\meter} in the $L=\SI{21}{\centi\meter}$ trace of \mysubfigref{phase_length}{a}.
Longer fibre lengths will strip out these HOMs, so no further peaks are observed; even for short lengths this beating is eliminated when the coupling is fully optimised.
Full optimisation of the coupling was not performed here, and so these additional HOMs remain visible at short $\LHCF$.

The linear relationship derived in \myeqnref{beating} is closely related to the group delay difference bewtween the two modes.
We denote this quantity $\tau=1/(\LHCF\Delta\omega)$, and it can be calculated from the gradient of the slope in \mysubfigref{phase_length}{b} to be $\tau=\SI{0.11+-0.03}{\pico\second\per\meter}$.

\begin{figure}[htbp]
  \centering
  \includegraphics[width=8cm]{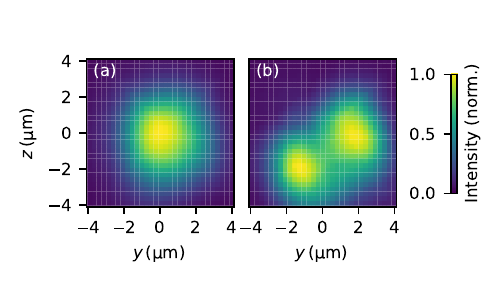}
  \caption{The $S^2$ measurements of the fundamental (a) and first order (b) modes of light propagating through the HCF (both normalised to a maximum of one). Light is coupled into the HCF by a SMF-GIF-HCF interconnect similar to that used in the analysed device.}
  \label{S2}
\end{figure}

The first HOM is further characterised by spatially and spectrally resolved ($S^2$) imaging~\cite{Nicholson2008} of light exiting the HCF, when coupled into by the same type of SMF-GIF-HCF splice that was used to build the low-loss vapour cell.
Spatial measurement of the light was achieved by scanning a 780HP fibre across the near field, with collimation and magnification performed by a microscope objective.
The light was directed to an Ocean Optics HR4000 for spectral measurement.
The fundamental mode and first detected HOM are shown in \myfigref{S2}, with a calculated group delay difference of $\tau=\SI{0.13+-0.02}{\pico\second\per\meter}$. We measure a power difference of \SI{1}{\percent} between the two modes, commensurate with the earlier result. The HOM does indeed look like an $\text{LP}_{1,1}$ mode.

\begin{figure*}[htbp]
  \centering
  \includegraphics[width=130mm]{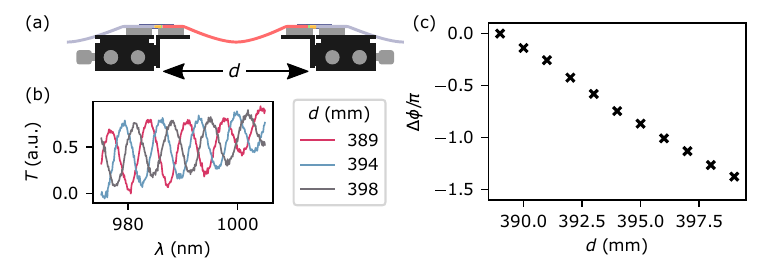}
  \caption{Changing phase of beating across the device. (a) Schematic of the experiment in which distance between fibre ends ($d$) is varied, allowing the fibre to droop. (b) Subset of data used to analyse the beating, showing varying phase of transmission ($T$) at different $d$. (c) Fourier analysis of the beating, showing that a full $\pi$ phase shift can be achieved.}
\label{phase_phase}
\end{figure*}

Loss due to the beating between modes can be mitigated by introducing a perturbation into the fibre such that interference at the desired wavelength is constructive. We do this by mounting the device on the alignment stages and bringing the ends of the fibre closer together, so that the fibre droops under its own weight as shown in \mysubfigref{phase_phase}{a}.
Varying distance between the HCF ends ($d$) allows control of the phase matching, and hence allows the optimisation of transmission at any given wavelength.
In \myfigref{phase_phase} we show the change in the phase of the transmission ($\Delta\phi$) relative to an arbitrary zero.
A change in $d$ of only a few centimeters can induce a full $\pi$ phase change, allowing the optimisation of the transmission at any wavelength in the transmission band.

\section{Conclusion}

We have demonstrated an encapsulated, fibre-integrated cell, consisting of a segment of HCF, with interconnects to SMF on either side.
The cell is intended to be used as a component in photonic quantum systems which have stringent requirements for loss.
These are satisfied here by means of a GIF segment, which allows for highly efficient mode matching between the SMF and HCF.
The interconnects encapsulate the core of the HCF, forming an enclosed cell, which, once filled with an atomic vapour, can be used as part of a memory or multiplexer for photonic quantum devices.
We demonstrated the efficiency of the device, which exhibits a total insertion loss of only \SI{0.6+-0.2}{\decibel} at \SI{780}{\nano\meter} wavelength, which is critical for storage and retrieval of nonclassical states of light in Rb-based quantum memories.
In principle there is nothing to prevent similar low-loss cells at any optical wavelength.

We have characterised the modal content supported by the HCF cell, both by observing beating between modes across various lengths of HCF and by means of an $S^2$ measurement of light exiting the HCF.
This has confirmed that the light propagates through the HCF predominantly in the fundamental mode, allowing robust SMF-HCF coupling at both interfaces and high efficiency across the whole device.
Our observation of beating between the fundamental and first higher-order mode enabled the transmission at any wavelength within the HCF transmission band to be optimised by allowing the HCF to bend under its own weight.
We anticipate that refinements to the interconnect fabrication process will enable further reduction of the interconnect loss, by improving fibre concentricity in the support capillary and fully optimising the GIF length (for example by polishing the GIF to more precisely control the mode profile on exit).

Future work will turn to the prospect of filling the device with rubidium vapour and the realisation of high-efficiency quantum memory inside a fibre-integrated cell.
Novel filling techniques will be required to fully realise an in-fibre atomic vapour cell, with possibilities including end-loading rubidium before connectorisation or side loading a complete cell through holes laser machined through the jacket of the HCF.

%

\printbibliography

\end{document}